\documentclass[twocolumn,aps,prb,superscriptaddress,10pt,showpacs]{revtex4-1}	% Article class of KOMA-script with 11pt font and a4 format
\usepackage{amsmath,amsfonts,amsthm}			% Math packages
\usepackage{relsize}
\usepackage{adjustbox,lipsum}
\usepackage{float}
\usepackage[caption=false]{subfig}
\usepackage{natbib}
\usepackage[ dvipdfm, 
                 bookmarks,
                 bookmarksopen = true,
                 bookmarksnumbered = true,
                 breaklinks = true,
                 linktocpage,
                 colorlinks = true,
                 linkcolor = blue,
                 urlcolor  = blue,
                 citecolor = blue,
                 anchorcolor = green,
                 hyperindex = true,
                 hyperfigures
                 ]{hyperref}
\usepackage{url}    %base package for URL support
\usepackage{epsfig}
  \makeatletter
    \renewcommand\@make@capt@title[2]{%
     \@ifx@empty\float@link{\@firstofone}{\expandafter\href\expandafter{\float@link}}%
      {\textbf{#1}}\@caption@fignum@sep#2\quad}%
    \makeatother
   
\makeatletter
\renewcommand{\fnum@figure}{\textbf{Figure~\thefigure}}
\begin{document}
\title{A self-consistent hybrid functional for condensed systems}
\author{Jonathan H Skone}
\affiliation{Department of Chemistry, University of California Davis, One Shields Ave., Davis, CA 95616}
\author{Marco Govoni}
\affiliation{Institute for Molecular Engineering, University of Chicago, 5801 South Ellis Avenue, Chicago, IL 60637}
\author{Giulia Galli}
\email[Author to whom electronic correspondence should be addressed. Electronic address: ]{gagalli@uchicago.edu}
\affiliation{Institute for Molecular Engineering, University of Chicago, 5801 South Ellis Avenue, Chicago, IL 60637}
\date{\today}
\begin{abstract}A self-consistent scheme for determining the optimal fraction of exact-exchange for full-range hybrid functionals is presented and applied to the calculation of band gaps and dielectric constants of solids.
The exchange-correlation functional is defined in a similar manner to the PBE0 functional, but the mixing parameter is set equal to the inverse macroscopic dielectric function and it is determined self-consistently, by computing the optimal dielectric screening. We found excellent agreement with experiments for the properties of a broad class of systems, with band gaps ranging between 0.7 and 21.7 eV and dielectric constants wihin 1.23 and 15.9. We propose that the eigenvalues and eigenfunctions obtained with the present self-consistent hybrid scheme may be excellent inputs for $G_0W_0$ calculations.
\end{abstract} % end of abstract
\pacs{71.15.Mb, 31.15.E-, 77.22.-d}

\maketitle 
%
%%%%%%%%% Section I: Introduction %%%%%%%%%%%%%%
\section{ \textbf {Introduction}}
Density functional theory (DFT)\cite{Hohenberg:1964fz} continues to be a widely used theoretical methodology to describe both condensed matter and molecular systems. Its success is owed to the reasonably good accuracy in predicting numerous properties of broad classes of materials and molecules at a relatively low computational cost. In the Kohn-Sham (KS)\cite{Kohn:1965js} formulation of DFT the density dependent potential, Eq.~\eqref{equation:VKS}, is the sum of the Hartree $v_H$, the exchange-correlation $v_{xc}$, and the external potential of the nuclei $v_{ext}$: 
\begin{equation}
 v_{_{\mathrm{KS}}}(\mathbf{r}) =  v_H(\mathbf{r}) + v_{xc}(\mathbf{r})  + v_{ext}(\mathbf{r}).
\label{equation:VKS}
\end{equation}
The exact exchange-correlation potential is not known and it is approximated in various manners. To date, in the condensed matter physics community, the most widely used exchange-correlation functionals have been the local (LDA) and semilocal (GGA) ones\cite{Gross:DFTbook}. 
Another popular approximation makes use of so called hybrid functionals, defined by the sum of a local $v_{xc}$ and of a term proportional to the Hartree-Fock exact-exchange operator.\cite{Becke:1993fu} Within the generalized Kohn-Sham (GKS) formalism \cite{Seidl:1996vs} the total nonlocal potential $v_{_{\mathrm{GKS}}}(\mathbf{r,r'})$ is given by:
\begin{equation}
 v_{_{\mathrm{GKS}}}(\mathbf{r,r'}) =  v_H(\mathbf{r}) + v_{xc}(\mathbf{r,r'})  + v_{ext}(\mathbf{r})
\label{equation:GKS}
\end{equation}
where $v_{xc}$ is now fully nonlocal and can be expressed as
\begin{eqnarray}
 v_{xc}(\mathbf{r,r'}) &=& \beta v_x^{\mathrm{sr-ex}}(\mathbf{r,r';\omega}) + \alpha v_x^{\mathrm{lr-ex}}(\mathbf{r,r';\omega}) \label{equation:GKSxc} \\
            \nonumber                 && + (1-\beta)v_x^{\mathrm{sr}}(\mathbf{r;\omega}) +(1- \alpha )v_x^{\mathrm{lr}}(\mathbf{r;\omega})  + v_c(\mathbf{r}) \, .
\end{eqnarray}
In Eq.~\eqref{equation:GKSxc} $\alpha$ and $\beta$ are parameters that determine the amount of long-range and short-range exact-exchange, respectively. The long-range nonlocal potential $v_x^{\mathrm{lr-ex}}(\mathbf{r,r'};\omega)$ is defined as 
\begin{equation}
v_x^{\mathrm{lr-ex}}(\mathbf{r,r'};\omega) = -\sum\limits_{i=1}^{\mathrm{N_{occ}}} \phi_i(\mathbf{r})\phi^*_i(\mathbf{r'})\frac{\mathrm{erf}(\omega \mathbf{|r-r'|})}{|\mathbf{r-r'}|}
\label{equation:exact-exchange-lr}
\end{equation}
where $\omega$ is a parameter (separation length) and $\phi_i$ are single particle, occupied electronic orbitals. The short-range potential $v_x^{\mathrm{sr-ex}}(\mathbf{r,r'};\omega)$ is defined in a similar manner, with the complementary error function replacing the error function in Eq.~\eqref{equation:exact-exchange-lr}:
\begin{equation}
v_x^{\mathrm{sr-ex}}(\mathbf{r,r'};\omega) = -\sum\limits_{i=1}^{\mathrm{N_{occ}}} \phi_i(\mathbf{r})\phi^*_i(\mathbf{r'})\frac{\mathrm{erfc}(\omega \mathbf{|r-r'|})}{|\mathbf{r-r'}|} \, .
\label{equation:exact-exchange-sr}
\end{equation}
The Coulomb  potential is partitioned\footnote{Though we use, as an example, the error function to partition the Coulomb interaction into long- and short-range, other functions could be used such as the semiclassical Thomas-Fermi screening function.} as:
\begin{equation}
\frac{1}{|\mathbf{r-r'}|} = \frac{\mathrm{erfc}(\omega|\mathbf{r-r'}|)}{|\mathbf{r-r'}|} + \frac{\mathrm{erf}(\omega |\mathbf{r-r'}|)}{|\mathbf{r-r'}|} \, .
\end{equation} 
When $\alpha = \beta = 0$, one recovers the KS equations with a local or semilocal exchange-correlation potential. If $\alpha = \beta = 1$, one obtains the KS equations with exact-exchange potential. For short-range hybrid functionals $\alpha=0$, e.g. in HSE06\cite{Heyd:2006dc} where $\beta=0.25$ and $\omega=0.11$ bohr$^{-1}$ or in sX-LDA\cite{Bylander:1990we} where $\beta=1$ and the Thomas-Fermi screening factor is used instead of the error function. When $\alpha \ne 0$ the range-separated hybrid functional is long-ranged. 
Examples of long-range hybrid functionals include the empirical CAM-B3LYP functional,\cite{Yanai:2004we} where $\alpha=0.65, \beta=0.19, \omega=0.33$ bohr$^{-1}$, as well as LC-$\omega$PBE,\cite{Weintraub:2009ub} where $\alpha=1$, $\beta=0$, and $\omega=0.4$ bohr$^{-1}$.
When $\alpha = \beta$, a full-range hybrid functional is obtained and $\alpha$ determines the fraction of exact-exchange entering the definition of the potential:  
\begin{equation}
 v_{xc}(\mathbf{r,r'}) =  \alpha v_{x}^{\mathrm{ex}}(\mathbf{r,r'}) + (1-\alpha)v_{x}(\mathbf{r}) + v_{c}(\mathbf{r}) \, ,
\label{equation:single-param-hybrid}
\end{equation}
where $v_{x}^{\mathrm{ex}}(\mathbf{r,r'})$ corresponds to the sum of the exact-exchange terms of Eq.~\eqref{equation:exact-exchange-lr} and Eq.~\eqref{equation:exact-exchange-sr} and similarly $v_{x}(\mathbf{r})$ corresponds to the sum of the local exchange terms in Eq.~\eqref{equation:GKSxc}.

An example of a full-range hybrid is PBE0\cite{Adamo:1999hv}, where $\alpha=0.25$.
Hybrid functionals have been regularly used to describe molecules,\cite{Bickelhaupt} but their application to condensed matter systems has been slower to realize due to the substantial increase in computational cost, with respect to local functionals, when using e.g. plane-wave basis sets.
However, in the last decade, due in part to several methodological advances,\cite{Wu:2009ub,Gygi:2009jn,Gygi:2013ic} hybrid functionals have been increasingly used to investigate a variety of periodic systems with plane-wave basis sets, and have been shown to surmount some of the shortcomings of local and semilocal functionals.\cite{Kummel:2008jo} 
The fraction of exact-exchange included in the potential greatly affects the calculated electronic structure and related quantities such as the static dielectric constant, the energy gap and equilibrium geometries. 
In most hybrid functionals used to date, the fraction of exact-exchange is kept fixed. 

Recently, several authors have suggested using $\alpha$ as an adjustable parameter to reproduce the experimental band gap of solids.\cite{Pozun:2011bk,Conesa:2012jx,Alkauskas:2008ej,Alkauskas:2011tc,Broqvist:2010gh}
For nonmetallic, condensed systems the screening of the long-range tail of the Coulomb interaction is proportional to the inverse of the static dielectric constant ($\epsilon_\infty^{-1}$) and it is thus intuitive to relate the parameter $\alpha$ in Eq.~\eqref{equation:single-param-hybrid} to $\epsilon_\infty^{-1}$.
One may also justify such relation by using many-body perturbation theory.\cite{Strinati:1988wu,Onida:2002vu} For example, in Hedin's equations,\cite{Hedin:1965hi} the exchange-correlation potential of Eq.~\eqref{equation:single-param-hybrid} is replaced by the self-energy $\Sigma$ which is a nonlocal and energy dependent operator. One of the most successful approximations to $\Sigma$ is the GW approximation,\cite{hybertsen:85gw} which has been extensively used in the last three decades to improve upon the single particle energies and wavefunctions obtained with local and semilocal DFT calculations.\cite{Aulbur:2000vz,Nguyen:2012iy,Pham:2013gs,Ping:2013cd,Bruneval:2014}
Since the GKS potential is not energy dependent, one may only draw a comparison between Eq.~\eqref{equation:single-param-hybrid} and Hedin's equation in the GW, static approximation, known as the static COulomb Hole plus Screened EXchange (COHSEX).\cite{Hedin:1965hi} The connection between hybrid functionals and the COHSEX approximation has been previously discussed\cite{Alkauskas:2011tc,Marques:2011fi,Moussa:2012ky}, and we consider it here in further detail. Within the COHSEX approximation the self-energy contains separable local and nonlocal potentials:
\begin{equation}
\Sigma(\mathbf{r},\mathbf{r'},\omega=0) = \Sigma_{COH}(\mathbf{r},\mathbf{r'}) + \Sigma_{SEX}(\mathbf{r},\mathbf{r'}) \, ,
\label{eq:selfenergy}
\end{equation}
where the local $\Sigma_{COH}$ represents the Coulomb-hole (COH) interaction and the nonlocal $\Sigma_{SEX}$ is the statically screened exchange (SEX): 
\begin{eqnarray}
\Sigma_{COH}(\mathbf{r},\mathbf{r'}) &=& -\frac{1}{2} \delta(\mathbf{r}-\mathbf{r'})\left[v(\mathbf{r},\mathbf{r'})- W(\mathbf{r},\mathbf{r'})\right] \label{eq:coh}
\\
\Sigma_{SEX}(\mathbf{r},\mathbf{r'}) &=& -\sum_{i=1}^{\mathrm{N_{occ}}} \phi_i(\mathbf{r})\phi^{\ast}_i(\mathbf{r'}) W(\mathbf{r},\mathbf{r'})\, .
\label{eq:sex}
\end{eqnarray}
In Eq.~\eqref{eq:coh} and~\eqref{eq:sex} the screened Coulomb potential $W$ is given by 
\begin{equation}
W(\mathbf{r},\mathbf{r'}) = \int d{\mathbf{r''}}  \epsilon^{-1} (\mathbf{r},\mathbf{r''}) v(\mathbf{r''},\mathbf{r'}),
\label{eq:w}
\end{equation}
where $\epsilon^{-1}$ is the dielectric response function and $v$ is the bare Coulomb potential.
If we approximate the inverse microscopic dielectric function $\epsilon^{-1}$ by the inverse macroscopic dielectric constant $\epsilon^{-1}_{\infty}$, thereby neglecting the  microscopic components of the dielectric screening, we obtain:
\begin{equation}
W(\mathbf{r},\mathbf{r'}) \approx \frac{1}{\epsilon_{\infty}} v(\mathbf{r},\mathbf{r'}) \,.
\label{eq:wstat}
\end{equation}
Inserting Eq.~\eqref{eq:wstat} in Eq.~\eqref{eq:coh} and~\eqref{eq:sex} yields the following expressions for COH and SEX:
\begin{eqnarray}
     \Sigma_{COH}(\mathbf{r},\mathbf{r'}) &\approx&  -(1-\epsilon^{-1}_{\infty})\frac{1}{2} \delta(\mathbf{r}-\mathbf{r'})v(\mathbf{r},\mathbf{r'}) \label{eq:COH-simp}\\
                   \Sigma_{SEX}(\mathbf{r},\mathbf{r'})  &\approx& -\epsilon^{-1}_{\infty} \sum_{i=1}^{\mathrm{N_{occ}}} \phi_i(\mathbf{r})\phi^{\ast}_i(\mathbf{r'}) {v(\mathbf{r},\mathbf{r'})}.
\label{eq:SEX-simp}
\end{eqnarray}

We may now compare the exchange-correlation potential of Eq.~\eqref{equation:single-param-hybrid} and the electron self-energy of Eq.~\eqref{eq:selfenergy} using the simplified expressions for COH and SEX given by Eq.~\eqref{eq:COH-simp} and Eq.~\eqref{eq:SEX-simp}. 
 If $\alpha=\epsilon^{-1}_{\infty}$ is chosen, the prefactors of the local and nonlocal exchange potentials in Eq.~\eqref{equation:single-param-hybrid} are the same as those of the corresponding local and nonlocal self-energies in Eq.~\eqref{eq:COH-simp} and Eq.~\eqref{eq:SEX-simp}, respectively. Hence through simplifications of the many-body self-energy, Eq.~\eqref{eq:selfenergy}, we obtain $\alpha=1/\epsilon_{\infty}$. 
We also note that the equivalence between Eq.\eqref{equation:single-param-hybrid} and Eqs.~\eqref{eq:COH-simp}-\eqref{eq:SEX-simp} holds exactly for the nonlocal terms where the exact-exchange is present in both; the local operator arising from the COH part is expressed in Eq.~\eqref{equation:single-param-hybrid} using a local/semilocal form.  
A similar proportionality between $\alpha$ and $\epsilon^{-1}_{\infty}$ was derived from many-body perturbation theory to study the polarizability of semiconductors in the framework of time-dependent DFT, where $\alpha$ was used to statically screen the long-range contribution to the exchange-correlation kernel in the polarizability, but without introducing a nonlocal potential in the Hamiltonian.\cite{Botti:2004}

The use of the static electronic dielectric constant ($\epsilon_\infty$) to represent the effective screening of the exact-exchange potential in nonmetallic condensed systems has been previously suggested by several authors.\cite{Shimazaki:2008tr,Marques:2011fi,RefaelyAbramson:2013bv}
Marques \textit{et al.}\cite{Marques:2011fi} evaluated $\epsilon_\infty$ at the semilocal PBE level of theory and set $\alpha={1/\epsilon}^{\text{PBE}}_{{\infty} }$ using a full-range hybrid functional. Building upon previous work on range-separated hybrid functionals for molecules, Refaely-Abramson \textit{et al.}\cite{RefaelyAbramson:2013bv} determined the static dielectric constant from the full dielectric response function in the random phase approximation (RPA). The results of both Ref.~\onlinecite{Marques:2011fi} and~\onlinecite{RefaelyAbramson:2013bv} showed a considerable improvement in computed electronic energy gaps, over semilocal and hybrid functionals. Despite properly describing the correct long-range asymptotic limit the accuracy of the prefactors used in Ref.~\onlinecite{Marques:2011fi} and Ref.~\onlinecite{RefaelyAbramson:2013bv}, may be affected by the level of theory (PBE) or the approximations employed for the evaluation of the polarizability (RPA). 

Using both a full-range and a short-range screened hybrid functional, Shimazaki \textit{et al.}\cite{Shimazaki:2009hx,Shimazaki:2010dn} self-consistently evaluated $\alpha = \epsilon^{-1}_\infty$ by using a Penn model for the static dielectric constant.\footnote{In the Penn model \cite{Penn:1962} the static dielectric constant is approximated as $\epsilon_{\infty} \approx 1+\left(\frac{h\omega_p}{E_g}\right)^2$ where $\omega_p$ is the plasmon frequency and $E_g$ is the energy gap.}
Koller \textit{et al.}\cite{Koller:2013ht} also reported a self-consistent short-range hybrid functional with the short-range mixing parameter dependent on the static dielectric constant. The latter was evaluated without including the density response to the perturbing external electric field (no local-field effects); an empirical fit was utilized to set the relation between $\alpha$ and $\epsilon_{\infty}$, resulting in considerable errors ($\sim30$\%) in the computed macroscopic dielectric constants. 
The self-consistent hybrid implementations of Ref.~\onlinecite{Shimazaki:2009hx} and~\onlinecite{Koller:2013ht} used approximate methods for the polarizability, which may have affected the overall accuracy of the procedure.

In this work we present a full range, non empirical hybrid functional where the mixing parameter $\alpha$ is determined self-consistently from the evaluation of the inverse static electronic dielectric constant $\epsilon^{-1}_\infty$. The latter is computed by including the full response of the electronic density to the perturbing external electric field, i.e. local-field effects are included, which are important to obtain accurate results. 
We computed the dielectric constants, electronic gaps and several lattice constants of a broad class of solids and found results in considerably better agreement with experiments than those obtained with semilocal and the PBE0 hybrid functional.

The rest of the paper is organized as follows. Section II describes the methodology along with the computational details. Section III presents results obtained using a self-consistent (sc) hybrid. Section IV summarizes the present self-consistent hybrid scheme and concludes with future directions to explore.

%%%%%%%%%%%%%% Section II Methods %%%%%%%%%%%%%%%%%%%
\section{ \bf {Methods}}
%%%%%%%%%%%%%%%%%%%%%%%%%%%%%%%%%%%%%%%%%%%%%%%%%%%%%
\subsection{\bf {Self-consistent hybrid mixing scheme (sc-hybrid)}}
\begin{figure}[]
\centering
\scalebox{0.3}{\includegraphics[trim= 0mm 0mm 0mm 0mm,clip, angle=90]{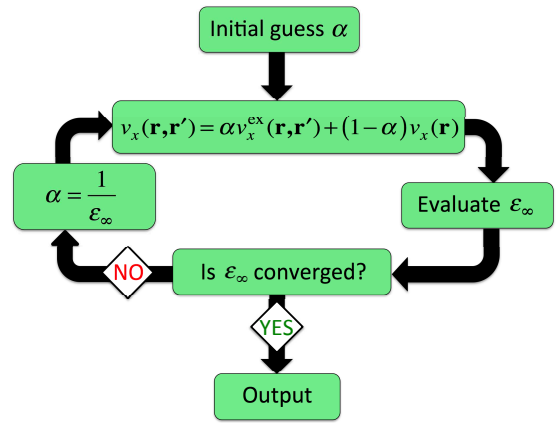}}
\caption{(Color online) Diagram of the self-consistent hybrid scheme. The potential used in the solution of the generalized Kohn-Sham equation is defined in Eq.~\eqref{equation:single-param-hybrid}. $\epsilon_{\infty}$ is the static dielectric constant.}
\label{fig:sc-hybrid-diagram}
\end{figure}

The self-consistent cycle used to determine the sc-hybrid functional proposed in this work is shown in Fig.~\ref{fig:sc-hybrid-diagram}. 
The self-consistency loop is started with an initial guess for $\alpha$, which is bound to range from 0 to 1; $\alpha$ determines the amount of exact-exchange $v_x^{\mathrm{ex}}(\mathbf{r,r'})$ included in the exchange-correlation potential expression of Eq.~\eqref{equation:single-param-hybrid}.  In this work we used the GGA exchange and correlation functional proposed by Perdew, Burke, and Ernzerhof (PBE),\cite{Perdew:1996ug} hence in Fig.~\ref{fig:sc-hybrid-diagram} $v_x(\mathbf{r})$ denotes the PBE exchange functional. 
Once the hybrid exchange potential is defined, $\epsilon_{\infty}$ is computed self-consistently using the procedure outlined in Section II.B and convergence is assessed by comparing $\epsilon_\infty$ evaluated in subsequent cycles. 

\begin{figure}[]
\scalebox{0.8}{\includegraphics[width=0.65\textwidth,trim= 0mm 0mm 0mm 0mm,clip]{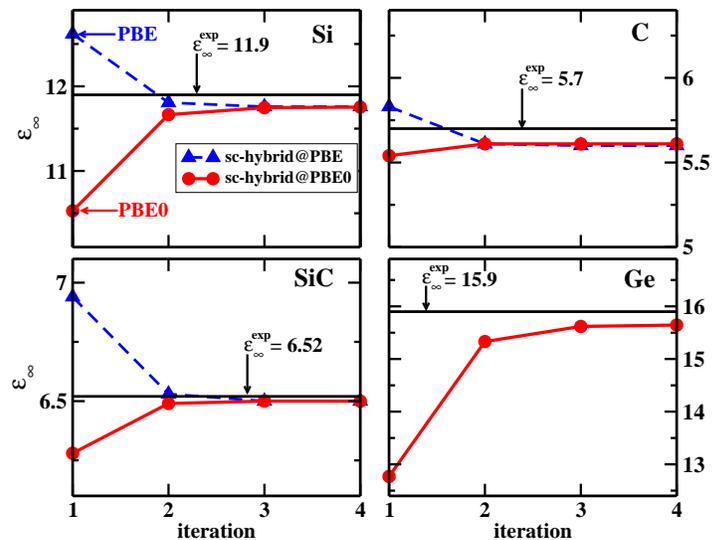}}
\caption{(Color online) Convergence of the value of the static dielectric constant $\varepsilon_{\infty}$ in the sc-hybrid scheme is shown for four prototypical semiconductors, Si, C, SiC, and Ge. The blue dashed line and triangles indicate the iterative procedure that starts with no inclusion of exact-exchange: $\alpha$=0 (sc-hybrid@PBE); the red solid line and circles correspond to the iterative procedure started with a quarter of exact-exchange $\alpha$=0.25 (sc-hybrid@PBE0). The blue and red arrows shown in the first panel indicate the PBE and PBE0 values of $\epsilon_{{\infty}}$. The solid black lines represent the value of the experimental macroscopic dielectric constant.}\label{fig:epsilon-convg-all}
 \end{figure}

As an initial guess for $\alpha$ we used both the value that reproduces the semilocal-only PBE limit ($\alpha=0$) and the value $\alpha=0.25$ corresponding to the global hybrid PBE0.
Fig.~\ref{fig:epsilon-convg-all} illustrates how the self-consistent procedure of Fig.~\ref{fig:sc-hybrid-diagram} leads to the same converged electronic dielectric constant regardless of the initial value of $\alpha$, either PBE (sc-hybrid@PBE, blue dashed line and triangles), or PBE0 (sc-hybrid@PBE0, red solid line and circles). Generally only three to four iterations are required to reach convergence,\footnote{Convergence of the sc-hybrid scheme is achieved when $\epsilon_{\infty} < 0.01 $. In the present scheme $\epsilon_{\infty}$ is only evaluated at the end of each converged SCF calculation; in principle one could instead evaluate $\epsilon_{\infty}$ at each step of the SCF procedure, in an attempt to reduce the overall number of SCF iterations, but this would be prohibitively expensive.} with the only notable exceptions being the antiferromagnetic transition metal oxides--CoO, MnO, and NiO, which respectively required 5, 5, and 9 iterations to reach convergence.

%%%%%%%%%%%%%%%%%%%%%%%%%%%%%%%%%%%%%%%%%%%%%%%%%%%%%%%%%%%%%%%
\subsection{ \bf {Evaluation of the static dielectric constant}}
The static dielectric constant is the central quantity in the sc-hybrid scheme and its accurate computation is critical for the performance of our approach. It is therefore useful to briefly recall the techniques and the levels of approximation that are usually employed in evaluating $\epsilon_\infty$.

We consider the dielectric response of a system subject to a macroscopic electric field $\mathbf{E}_{ext}$, where the total potential acting on the system $v_{tot}$ includes both the perturbing macroscopic potential $v_{macro}= e \mathbf{r}\cdot\mathbf{E}_{ext}$, and the self-consistent generalized Kohn-Sham electronic potential $v_{_{\textrm{GKS}}}$:
\begin{equation}
v_{tot} = v_{_{\textrm{GKS}}} + v_{macro} \,.
\label{eq:vtot}
\end{equation}
The dielectric response to an external field may be computed using finite field methods, e.g. the Berry phase technique (known as the modern theory of polarization),\cite{KingSmith:1993wg,Resta:1994cr} or first order perturbation theory, which is our method of choice in the present work.
Within linear response, both Density Functional Perturbation Theory (DFPT)\cite{Baroni:2001ia} and the coupled perturbed Kohn-Sham (CPKS)\cite{Rerat:2008th,Johnson:1993db} equations (the coupled-perturbed Hartree-Fock method (CPHF)\cite{Pople:1979hm,Hurst:1988gp,Orlando:2009fu} extended to DFT) have been commonly employed to compute the macroscopic dielectric constants of solids. In this work we computed the dielectric constants using the CPKS method as implemented in CRYSTAL09,\cite{Dovesi:2005eq} where the perturbed KS orbitals are obtained using the potential of Eq.~\eqref{eq:vtot}. 
The $v_{_{\textrm{GKS}}}$ potential implicitly depends on the applied electric field through the perturbed charge density and orbitals.
The perturbation of the equilibrium charge density caused by the presence of the external field is related to $v_{macro}$ by the reducible polarizability $\chi$:
\begin{equation}
n_{ind} (\mathbf{r}) = \int  \chi(\mathbf{r},\mathbf{r}^\prime) v_{macro}(\mathbf{r}^\prime) d\mathbf{r^\prime}.
\label{eq:nind}
\end{equation}
$\chi$ is a nonlocal operator that describes the many-body polarization effects of the interacting electron gas. The polarizability $\chi$ may include retardation effects, giving rise to a frequency dependence of the dielectric tensor. Such dependence is not considered in the present work since we focus on the evaluation of the static dielectric screening. 
The static dielectric tensor $\epsilon^{-1}_{ij}$ can be expressed in terms of $\chi$\cite{Ehrenreich:1966vz}:
\begin{equation}
\epsilon^{-1}_{ij} = \delta_{ij} + \frac{4\pi e^2}{\Omega} \int d\mathbf{r} \int d\mathbf{r}^\prime r_i \chi(\mathbf{r},\mathbf{r}^\prime) r^\prime_j 
\label{eq:epsilonwe}
\end{equation}
where $i,j$ denote Cartesian components and $\Omega$ is the volume of the cell. 
This result can be derived by relating the external electric field $\mathbf{E}_{ext}$ to the total electric field $\mathbf{E}=\mathbf{E}_{ext}-4\pi\mathbf{P}$ and by computing the induced polarization field $\mathbf{P}$ by integrating the induced charge density
\begin{equation}
\mathbf{P} = \frac{-e}{\Omega} \int   n_{ind} (\mathbf{r}) \mathbf{r} d\mathbf{r} \,.
\label{eq:pola}
\end{equation}
The approximations adopted in the computation of the static dielectric constant arise from the approximation chosen for $\chi$ in Eq.~\eqref{eq:epsilonwe}: 
\begin{eqnarray}
\nonumber \chi &=& \chi_0 + \chi_0 \frac{\delta v_{_{\textrm{GKS}}}}{\delta n} \chi_0 +  \chi_0 \frac{\delta v_{_{\textrm{GKS}}}}{\delta n} \chi_0 \frac{\delta v_{_{\textrm{GKS}}}}{\delta n} \chi_0 + \cdots \\
&=& \chi_0 + \chi_0 \frac{\delta v_{_{\textrm{GKS}}}}{\delta n} \chi 
\label{eq:dysonchi}
\end{eqnarray}
where $\chi_0$ is the irreducible polarizability\cite{Hanke:1978tj}. The reducible and irreducible polarizabilities are also called interacting and non-interacting density-density response functions, respectively.\cite{fetterwalecka} The difference between $\chi$ and $\chi_0$ is given by so called local-field effects\cite{Alder:1962,Wiser:1963} defined by the functional derivative $\frac{\delta v_{_{\textrm{GKS}}}}{\delta n}$ which is the sum of the functional derivative of the Hartree and exchange-correlation potential with respect to the density:
\begin{equation}
\frac{\delta v_{_{\textrm{GKS}}}}{\delta n} =  \frac{\delta v_{H}}{\delta n} +  \frac{\delta v_{xc}}{\delta n} \, ,
\label{eq:vpfxc}
\end{equation}
where $\frac{\delta v_{xc}}{\delta n} \equiv f_{xc}$. If local-fields are neglected (NLF, i.e. no local-fields), $\frac{\delta v_{_{\textrm{GKS}}}}{\delta n}=0$ and the polarizability is equal to the irreducible polarizability 
\begin{equation}
\chi^{NLF} = \chi_0 \, .
\label{eq:chi-NLF}
\end{equation}
$\chi_0$ can then be computed in the independent particle approximation (IPA), which assumes the electron-hole (e-h) interactions are negligible.\cite{gross:1991} 
This approximation is formally equivalent to the one adopted for the calculation of $\epsilon_\infty$ by Koller \textit{et al.}\cite{Koller:2013ht}, who used the Fermi's golden rule to compute the frequency-dependent imaginary part of the dielectric screening, and the Kramers-Kronig relation to derive the static dielectric constant. \\
If only the Hartree term is included in Eq.~\eqref{eq:vpfxc}, and the derivative of the exchange-correlation potential is set to zero ($f_{xc}=0$) one obtains the random phase approximation (RPA) for the polarizability:
\begin{equation}
\chi^{RPA} = [1-\chi_0 v ]^{-1} \chi_0 \, .
\label{eq:chirpa}
\end{equation}
If both the Hartree and the exchange-correlation terms are included in Eq.~\eqref{eq:vpfxc}, one obtains $\textrm{RPA}+f_{xc}$.\cite{Paier:2008ur}
In linear response calculations of the dielectric screening with nonlocal operators in the KS potential, the functional derivative of $v_{xc}$ with respect to the density is usually neglected; the resulting, approximate $f_{xc}$ is denoted as $f_{xc-l}$. 
If the functional derivative of the nonlocal operator is instead included, $f_{xc}$ is denoted as $f_{xc-nl}$. 
We note that when using local/semilocal functionals, the exchange-correlation potential entering $f_{xc}$ depends explicitly on the density and its functional derivative may be readily evaluated; its inclusion in the calculation of the polarizability for some semiconductors and insulators was previously observed to be negligible.\cite{Paier:2008ur}
However, nonlocal exchange-correlation potentials, e.g. derived from hybrid functionals, depend implicitly on the density through the KS orbitals and their functional derivative is not straightforward to compute. Within the CPKS method this difficulty is overcome by calculating explicitly the perturbed orbitals and using them to evaluate the linear variation of the exact-exchange with respect to the single particle wavefunctions; hence within the CPKS scheme local-field effects are easily included.\footnote{The explicit calculation of the perturbed orbitals comes at the expense of evaluating unoccupied KS orbitals that enter the expression of the functional derivatives. Within DFPT applied to semilocal functionals the calculations of empty orbitals is avoided by utilizing projection techniques.\cite{Baroni:2001ia}}

The importance of including nonlocal contributions to $f_{xc}$ in the calculation of band-gaps of some semiconductors and insulators was pointed out by Paier \textit{et al.} \cite{Paier:2008ur}, following the suggestion of Bruneval \textit{et al.}\cite{Bruneval:2005vn}; these authors derived $f_{xc}$ from many-body perturbation theory and related it to the inclusion of e-h interactions in the many-body calculations of $\chi_0$, beyond the IPA. 

Finally we note that the CPKS scheme adopted here is efficient when used in conjunction with moderate size basis sets, e.g. the Gaussian basis sets we used with CRYSTAL09. However this would not be as practical when plane-wave basis sets are employed. Within a plane-wave pseudopotential approach with hybrid functionals, one may for example evaluate the dielectric constant by applying the modern theory of polarization and computing derivatives with respect to the applied field by finite differences. In this way all local-field effects are automatically included.\cite{KingSmith:1993wg,Resta:1994cr}

%%%%%%%%%%%%%%%%%%%%%%%%%%%%%%%%%%%%%%%%%
\subsection{ \bf {Computational details}}
All hybrid functional calculations were carried out within an all-electron approach using the CRYSTAL09\cite{Dovesi:2005eq} electronic structure package. We thus avoided possible inconsistencies generated by the use of pseudopotentials derived within PBE, for hybrid functional calculations.
We used Gaussian basis sets modified starting from Alhrich's def2-TZVPP molecular basis,\cite{Weigend:2005dh} with the only exception of the rare gases Ne and Ar basis sets, which were modified starting from the def2-QZVPD set.\cite{Rappoport:2010} The highly contracted core shells were not modified, while the valence shells were modified, when necessary, to avoid possible linear dependencies caused by the use of  diffuse functions, which are utilized in the case of molecules to represent the tail of the wavefunctions in the vacuum region. In particular, we constrained the most diffuse exponents to be larger than or equal to 0.09 bohr$^{-2}$.
In most cases, we kept the size of the valence shell basis set to be the same as that of the uncontracted original def2 sets by augmenting the truncated basis sets accordingly. The Gaussian basis functions added to the original set were chosen so as to generate a basis set as even-tempered as possible. 
The orbital exponents of the augmented uncontracted valence shells were variationally optimized for all solids in the current study, using the GGA functional PBE. 
 
We note that a much denser {\it k}-point mesh is required for the convergence of the electronic dielectric constants than for the ground state energies and electronic energy gaps (see Supplementary Material).
In all calculations carried out with the sc-hybrid scheme we used the {\it k}-point mesh required to converge $\epsilon_\infty$.

We also carried out plane-wave calculations at the GGA level of theory using the Quantum-ESPRESSO plane-wave pseudopotential package\cite{Giannozzi:2009hx} to compare with the results of CRYSTAL09. We employed both the projector-augmented wave function (PAW) pseudopotentials and norm-conserving pseudopotentials, which were either generated using the ATOMPAW program\cite{Tackett:2001em} or obtained from the Quantum-ESPRESSO pseudopotential library.\cite{QEpsp} 
For the transition metal atoms, unless otherwise noted, the $(n-1)$s and $(n-1)$p electrons, where $n$ is the highest principle quantum number, were always included in the valence. 
Additional information on the pseudopotentials used in plane-wave calculations, $k$-point convergence for the polarizabilities, and a comparison of the plane-wave and Gaussian basis sets is provided in the Supplemental Material.

With the exception of the lattice optimizations, all calculations were performed at the experimental geometry and $T=0K$. 

%%%%%%%%%%%%%%%% Results and Discussion %%%%%%%%%%%%%%%%%%
\section{ \bf {Results and Discussion} }
%%%%%%%%%%%%%%%%%%%%%%%%%%%%%%%%%%%%%%%%%%%%%%%%%%%%%%%%
\subsection{{\bf Static electronic dielectric constant }}
The static electronic dielectric constant ($\epsilon_{\infty}$) of several crystalline materials, was evaluated with PBE, PBE0, the fixed-$\alpha$ hybrid functionals ($\alpha={1/\epsilon}^{\text{PBE}}_{{\infty} }$  and $\alpha={1/\epsilon}^{\text{PBE0}}_{{\infty} }$), and the self-consistent version (sc-hybrid). In Table~\ref{table:epsilinf-comp-tab} results are shown for 24 solids, which cover a broad range of static dielectric constants (1.23--15.9) and band gaps (0.7--21.7 eV). In the case of non-cubic systems, we report the average of the trace of the dielectric tensor. 
Results obtained with the semilocal PBE and the hybrid PBE0 functionals exhibit the poorest agreement with experiment, with the PBE error being at least twice as large as those of other hybrid functionals. 
The closest agreement with experiment is obtained with the sc-hybrid, although using $\alpha = 1/\epsilon^{\text{PBE}}_\infty$ or $\alpha = 1/\epsilon^{\text{PBE0}}_\infty$ also yield satisfactory results.
We note that the absence of values for CoO and Ge in both the PBE and hybrid ($\alpha={1/\epsilon}^{\text{PBE}}_{{\infty} }$) columns is due to the fact that these systems turn out to be erroneously metallic when using semilocal functionals and thus $\epsilon^{\text{PBE}}_{{\infty}}$ cannot be evaluated.
\begin{table*}[] 
\caption{The electronic dielectric constant ($\epsilon_{\infty}$) determined using several levels of theory. The hybrid heading with $\alpha=1/\epsilon^{\text{PBE}}_{\infty}$ refers to a hybrid calculation using $\alpha={1/\epsilon}_{\infty}$ where the dielectric constant was evaluated at the PBE level of theory. Similarly the hybrid heading with $\alpha={1/\epsilon}^{\text{PBE0}}_{\infty}$ refers to a hybrid calculation where the dielectric constant was evaluated at the PBE0 level of theory. The sc-hybrid heading refers to hybrid calculations where the fraction of exact-exchange is self-consistently determined from the dielectric constant. All local-field effects are included in the evaluation of the dielectric constant so that all $\epsilon_{\infty}$ hybrid functional entries in the Table are at level of RPA+$f_{xc-nl}$. ME, MAE, MRE, and MARE are the mean, mean absolute, mean relative, and mean absolute relative error, respectively. The experimental geometry was used for each solid, with the structure/polytype indicated by the abbreviation in the second column: dC--Diamond Cubic; RS--Rock Salt cubic structure; ZB--Zinc Blende; M--Monoclinic; Ru--Rutile; WZ--Wurtzite; Cr--Corundum; XI--the XI proton ordered phase of ice; cF--fcc face centered cubic. Note that CoO, NiO, and MnO are magnetic with AFM-II magnetic ordering.} 
\label{table:epsilinf-comp-tab}
\scalebox{0.95}{
\begin{tabular}{lccccccc}
%\toprule
%\toprule
\hline
\hline
  &&  PBE & PBE0 & hybrid & hybrid   & sc-hybrid&   Exp.  \\  
  & Type& $\alpha=0$   & $\alpha=0.25$  & $\alpha={1/\epsilon}^{\text{PBE}}_{{\infty} }$ & $\alpha={1/\epsilon}^{\text{PBE0}}_{{\infty} }$ & $\alpha=1/$sc-${\epsilon}_{\infty}$ &  \\  
%\cmidrule{3-8} 
\hline
  Ge        &(dC)  &   -- & 12.77& --   & 15.33 & 15.65 &  15.9  \cite{VanVechten:1969tk}  \\ % done
  Si        &(dC)  & 12.62& 10.53& 11.81& 11.67 & 11.76 &  11.9  \cite{Cardona:2001}       \\
  AlP       &(ZB)  & 7.82 & 6.85 & 7.26 & 7.20  & 7.23  &  7.54  \cite{Cardona:2001}       \\
  SiC       &(ZB)  & 6.94 & 6.28 & 6.53 & 6.49  & 6.50  &  6.52  \cite{Cardona:2001}       \\
  TiO$_2$   &(Ru)  & 7.91 & 5.96 & 6.75 & 6.46  & 6.56  &  6.34  \cite{DeVore:1951tu}    \\
  NiO       &(RS)  & 16.98&4.74  & 9.20 & 5.12  & 5.49  &  5.76  \cite{Rao:1965ey}         \\
  C         &(dC)  & 5.83 & 5.54 & 5.61 & 5.61  & 5.61  &  5.70  \cite{Cardona:2001}       \\
  CoO       &(RS)  & --   & 4.52 & --   & 4.73  & 4.92  &  5.35  \cite{Rao:1965ey}         \\
  GaN       &(ZB)  & 5.78 & 5.00 & 5.19 & 5.12  & 5.14  &  5.30  \cite{Giehler:1995jg}     \\
  ZnS       &(ZB)  & 5.58 & 4.84 & 5.01 & 4.94  & 4.95  &  5.13  \cite{Cardona:2001}\\
  MnO       &(RS)  & 7.62 & 4.32 & 5.11 & 4.41  & 4.45  &  4.95 \cite{Plendl1969109}         \\
  WO$_3$    &(M)   & 5.46 & 4.60 & 4.79 & 4.68  & 4.72  &  4.81  \cite{Hutchins:2006vl}    \\
  BN        &(ZB)  & 4.59 & 4.37 & 4.40 & 4.39  & 4.40  &  4.50  \cite{Chen:1995}    \\
  HfO$_2$   &(M)   & 4.54 & 3.97 & 4.03 & 3.97  & 3.97  &  4.41  \cite{Balog:1977dj}    \\
  AlN       &(WZ)  & 4.54 & 4.15 & 4.18 & 4.16  & 4.16  &  4.18  \cite{Shokhovets:2003ik}    \\
  ZnO       &(WZ)  & 4.66 & 3.54 & 3.63 & 3.47  & 3.46  &  3.74  \cite{Ashkenov:2003kq}    \\
  Al$_2$O$_3$ &(Cr)& 3.27 & 3.07 & 3.03 & 3.01  & 3.01  &  3.10  \cite{French:2005de}    \\
  MgO       &(RS)  & 3.12 & 2.89 & 2.83 & 2.81  & 2.81  &  2.96  \cite{Lide:2010}          \\
  LiCl      &(RS)  & 2.96 & 2.82 & 2.78 & 2.77  & 2.77  &  2.70  \cite{VanVechten:1969tk}  \\
  NaCl      &(RS)  & 2.49 & 2.37 & 2.31 & 2.30  & 2.29  &  2.40  \cite{Bass:2009}  \\
  LiF       &(RS)  & 1.97 & 1.87 & 1.79 & 1.78  & 1.77  &  1.90  \cite{VanVechten:1969tk}  \\
  H$_2$O    &(XI)  & 1.80 & 1.73 & 1.66 & 1.65  & 1.65  &  1.72  \cite{Johari:1976tc}          \\
  Ar        &(cF)  & 1.74 & 1.70 & 1.66 & 1.66  & 1.66  & 1.66 \cite{Sinnock:1969tt}     \\
  Ne        &(cF)  & 1.28 & 1.24 & 1.21 &  1.21 & 1.21  & 1.23 \cite{Schulze:1974cd}        \\
% IN FUTURE IF UNSURE OF DIELECTRIC OF MATERIAL LOOK AT Van Vechten Phys Rev paper 1969
%\cmidrule{3-8} 
\hline
  ME      && 0.96 & -0.41 & 0.13  & -0.20  & -0.15  &    --  \\
  MAE     && 0.96 &  0.43 & 0.27  &  0.22  &  0.18  &    --  \\
  MRE (\%)  &&  18.5 &-5.1 & 1.4  & -3.8 &-3.1  &    --  \\
  MARE (\%) &&  18.5 & 6.2 & 5.6  &  4.5 & 4.0  &    --  \\
%\bottomrule
%\bottomrule
\hline
\hline
\end{tabular} }
\end{table*}

We also compared the sc-hybrid results with those of many-body perturbation theory in the $GW$ approximation for a subset of solids for which previous $G_0W_0$ and self-consistent $GW$ results were reported\cite{Shishkin:2007ji,Shishkin:2007PRL} (see Table~\ref{table:epsilinf-GW-comp-tab}). 
The $G_0W_0$ (RPA) calculations were carried out by evaluating the dielectric response in the random phase approximation, without updating the electronic wavefunctions. The sc$GW$ (e-h) calculations were carried out self-consistently, using a frequency independent (static approximation) dielectric response with a vertex correction in $W$ that--effectively--includes the electron-hole interaction (e-h). The dielectric constants evaluated with sc-hybrid have similar errors as those obtained with the sc$GW$ (e-h) approach. The agreement between sc-hybrid and sc$GW$ (e-h) results suggest that the inclusion of nonlocal-field effects in the evaluation of the $f_{xc}$, when computing $\epsilon_{\infty}$, may play a similar role as the inclusion of the vertex corrections in W, when carrying out $GW$ calculations. This interpretation is also supported by the comparison of sc-hybrid with $G_0W_0$(RPA) results, which show the poorest agreement with experiments. We recall that within RPA only the local-field effects coming from the Hartree potential are included. The case where no local-field effects are present in $\chi$ (Eq.~\eqref{eq:chi-NLF}) using the sc-hybrid scheme is shown in the column heading under sc-hybrid (NLF) in Table~\ref{table:epsilinf-GW-comp-tab}. In the NLF case, the error is about three times as large as the case where local-fields are included.

Overall the agreement between sc-hybrid and sc$GW$ (e-h) results suggests that the static approximation captures most of the screening in the bulk materials considered here, and that including the dynamical frequency dependence in the dielectric response is not critical to obtain accurate static dielectric constants.
\begin{table}[] 
\caption{The electronic dielectric constant evaluated using the sc-hybrid functional scheme is compared with results obtained from many body perturbation theory. The sc$GW$ (e-h) calculations\cite{Shishkin:2007PRL} used the HSE hybrid functional eigenvalues and orbitals as input and the $G_{0}W_{0}$ (RPA) calculations\cite{Shishkin:2007ji} used the PBE functional eigenvalues and orbitals as input. The sc-hybrid (NLF) heading refers to a hybrid calculation where the uncoupled-perturbed Kohn-Sham equation is used resulting in no inclusion of the local-field effects. The sc-hybrid (RPA $+ f_{xc-nl}$) heading refers to a sc-hybrid calculation where the polarizability includes all local-field effects.  ME, MAE, MRE, and MARE are the mean, mean absolute, mean relative, and mean absolute relative error, respectively. The experimental geometry was used for each solid, with the structure/polytype used indicated in Table~\ref{table:epsilinf-comp-tab}.}
\label{table:epsilinf-GW-comp-tab}
\begin{adjustbox}{max width=\columnwidth,max height=\textheight,keepaspectratio}
\scalebox{1.0}{
\begin{tabular}{lccccc}
%\toprule
%\toprule
\hline
\hline
              &\multicolumn{2}{c}{sc-hybrid }  & $G_0W_0$ \cite{Shishkin:2007ji} & sc$GW$ \cite{Shishkin:2007PRL} &  Exp.  \\  
              & (NLF) & (RPA $+ f_{xc-nl}$) & (RPA) & (e-h)  & \\  
%\cmidrule{2-6} 
\hline
  Ge          &15.05 &  15.65 & --  & 15.30 & 15.9 \\
  Si         &11.24 &  11.76 & 12.09& 11.40 & 11.9 \\
  AlP        &6.91 &  7.23  &  7.53 & 7.11 & 7.54 \\
  SiC        &5.96 &  6.50  &  6.56 & 6.48 & 6.52 \\
  C          &5.08 &  5.61  &  5.54 & 5.59 & 5.70 \\
  GaN        &4.49 &  5.14  &  5.68 & 5.35 & 5.30 \\
  ZnS        &4.54 &  4.95  &  5.62 & 5.15 & 5.13 \\
  BN         &3.93 &  4.40  &  4.30 & 4.43 & 4.50  \\
  ZnO        &2.89 &  3.46  &  5.12 & 3.78 & 3.74  \\
  MgO        &2.42 &  2.81  &  2.99 & 2.96 & 2.96  \\
  LiF        &-- &  1.77  &  1.96 & 1.98 & 1.90    \\
  Ar         &1.60 &  1.66  &  1.66 & 1.69 &1.66    \\
  Ne         &1.17 &  1.21  &  1.25 & 1.23 &1.23    \\
%\cmidrule{2-6} 
\hline
  ME        & -0.57  & -0.14 & 0.19  & -0.12 & -- \\
  MAE       & 0.57  & 0.14  & 0.25  & 0.15  & -- \\
  MRE (\%)  & -10.7  & -3.0  & 4.5   & -0.7  & -- \\
  MARE (\%) & 10.7  & 3.0   & 5.8   & 2.0   & -- \\
%\bottomrule
%\bottomrule
\hline
\hline
\end{tabular} }
\end{adjustbox}
\end{table}
%
%%%%%%%%%%%%%%% GaN and AlN table of epsilon for WZ and ZB phases %%%%%%%%%%%%%%%%%%
\begin{table*}[] 
\caption{The calculated electronic dielectric tensor components for the optically anisotropic wurtzite phases of ZnO, AlN, and GaN are shown, and compared with their respective experimental values.} 
\label{table:epsilinf-WZ-tab}
\scalebox{0.96}{
\begin{tabular}{lcccccccc}
%\toprule
%\toprule
\hline
\hline
 & \multicolumn{2}{c}{ZnO}& &\multicolumn{2}{c}{AlN} & &\multicolumn{2}{c}{GaN}   \\
               &   $\epsilon_{\perp}(\infty)$  &  $\epsilon_{\parallel}(\infty)$ &&   $\epsilon_{\perp}(\infty)$  &  $\epsilon_{\parallel}(\infty)$ &&   $\epsilon_{\perp}(\infty)$  &  $\epsilon_{\parallel}(\infty)$ \\
%\cmidrule{2-3} \cmidrule{5-6} \cmidrule{8-9} 
\hline
  PBE                 & 4.64    & 4.68 && 4.46  & 4.69   &&   5.55  & 5.72 \\ 
  PBE0                & 3.53    & 3.57 && 4.10  & 4.27   &&   4.86  & 5.00 \\ 
  sc-hybrid           & 3.45    & 3.48 && 4.10  & 4.27   &&   4.99  & 5.13 \\
  Experiment\cite{Shokhovets:2003ik,Ashkenov:2003kq} &   3.70 $\pm$0.01  & 3.78 $\pm$0.05  && 4.13 $\pm$0.02  & 4.27 $\pm$0.05 && 5.18 $\pm$0.02 & 5.31 $\pm$0.06 \\
%\bottomrule
%\bottomrule
\hline
\hline
\end{tabular} }
\end{table*}

To further evaluate the accuracy of the static electronic dielectric constants obtained with the sc-hybrid functional we compared the computed individual tensor components for the optically anisotropic wurtzite phases of GaN, AlN, and ZnO in Table~\ref{table:epsilinf-WZ-tab}. The agreement with experimental results is very good for each of the individual tensor components. 
%%%%%%%%%%%%%%%%%%%%%%%%%%%%%%%%%%%%%%%%%%%%%%%%%%%%%%%%%%
\subsection{{\bf Electronic energy gaps and band structure }}
\begin{table*}[] 
\caption{The Kohn-Sham (KS) energy gaps (eV) evaluated with the dielectric-dependent hybrid functionals are compared with the experimental electronic gaps for a wide range of materials. The experimental values correspond to either photoemission measurements or to optical measurements where the excitonic contributions were removed, with alumina the only exception (see text). The KS gaps were computed as the energy difference of the single particle energies of the conduction band minimum and the valence band maximum. The solids are listed in the order of largest to smallest experimental $\epsilon_{{\infty}}$. The hybrid heading with $\alpha={1/\epsilon}^{\text{PBE}}_{{\infty} }$ refers to a hybrid calculation using a fixed $\alpha$ with the dielectric constant evaluated at the PBE level of theory. Similarly the hybrid heading with $\alpha={1/\epsilon}^{\text{PBE0}}_{{\infty} }$ refers to a hybrid calculation with a fixed $\alpha$ and the dielectric constant evaluated at the PBE0 level of theory. The sc-hybrid  heading refers to the hybrid calculation where the fraction of exact-exchange is determined from the self-consistently from ${\epsilon}_{\infty}$.  ME, MAE, MRE, and MARE are the mean, mean absolute, mean relative, and mean absolute relative error, respectively. The experimental geometry was used in all calculations, with the structure/polytype indicated in the second column: dC--Diamond Cubic; RS--Rock Salt cubic structure; ZB--Zinc Blende; M--Monoclinic; Ru--Rutile; WZ--Wurtzite; Cr--Corundum; XI--the XI proton ordered phase of ice; cF--fcc face centered cubic. Note that CoO, NiO, and MnO are magnetic with AFM-II magnetic ordering.} 
\footnotetext[1]{the experimental value used here is the average of two reported values 6.1 and 6.4 eV.}
\label{table:Eg-comp-tab}
\scalebox{0.9}{
\begin{tabular}{llcccccc}
%\toprule
%\toprule
\hline
\hline
            &   &  PBE & PBE0 & hybrid & hybrid   & sc-hybrid&  Exp.  \\  
& Type & $\alpha=0$  & $\alpha=0.25$  & $\alpha={1/\epsilon}^{\text{PBE}}_{{\infty} }$ & $\alpha={1/\epsilon}^{\text{PBE0}}_{{\infty} }$ & $\alpha=1/$sc-${\epsilon}_{\infty}$  & \\  
%\cmidrule{3-8} 
\hline
  Ge      &(dC)    & 0.00 & 1.53  &  --  & 0.77 & 0.71 & 0.74 \cite{Kittel:2005}\\
  Si      &(dC)    & 0.62 & 1.75  & 0.96 & 1.03 & 0.99 & 1.17 \cite{Kittel:2005}\\
  AlP     &(ZB)    & 1.64 & 2.98  & 2.31 & 2.41 & 2.37 & 2.51 \cite{Monemar:1973jj}       \\
  SiC     &(ZB)    & 1.37 & 2.91  & 2.23 & 2.33 & 2.29 & 2.39 \cite{Choyke:1964fh}\\ 
  TiO$_2$ &(Ru)    & 1.81 & 3.92  & 2.83 & 3.18 & 3.05 & 3.3 \cite{Tezuka:1994cv} \\
  NiO     &(RS)    & 0.97 & 5.28  & 2.00 & 4.61 & 4.11 & 4.3 \cite{Sawatzky:1984jt}\\
  C       &(dC)    & 4.15 & 5.95  & 5.37 & 5.44 & 5.42 & 5.48 \cite{Clark:1964km} \\
  CoO     &(RS)    & 0.00 & 4.53  &  --  & 4.01 & 3.62 & 2.5 \cite{vanElp:1991ia} \\
  GaN     &(ZB)    & 1.88 & 3.68  & 3.10 & 3.30 & 3.26 & 3.29 \cite{RamirezFlores:1994ws}\\
  ZnS     &(ZB)    & 2.36 & 4.18  & 3.65 & 3.85 & 3.82 & 3.91 \cite{Kittel:2005}\\
  MnO     &(RS)    & 1.12 & 3.87  & 2.55 & 3.66 & 3.60 & 3.9  \cite{vanElp:1991uw} \\
  WO$_3$  &(M)    & 1.92 & 3.79  & 3.24 & 3.50 & 3.47 & 3.38 \cite{Meyer:2010cr} \\
  BN      &(ZB)    & 4.49 & 6.51  & 6.24 & 6.34 & 6.33 & 6.25  \cite{Levinshtein:2001}\footnotemark[1]\\
  HfO$_2$ &(M)     & 4.32 & 6.65  & 6.38 & 6.68 & 6.68 & 5.84  \cite{Sayan:2004gx}    \\
  AlN     &(WZ)    & 4.33 & 6.31  & 6.07 & 6.24 & 6.23 & 6.28  \cite{Roskovcova:1980gz}    \\
  ZnO     &(WZ)    & 1.07 & 3.41  & 3.06 & 3.73 & 3.78 & 3.44 \cite{Ozgur:2005it}\\
  Al$_2$O$_3$ &(Cr)&6.31  & 8.84  & 9.42 & 9.65 & 9.71 &  8.8 \cite{Innocenzi:1990bx}    \\
  MgO     &(RS)    & 4.80 & 7.25  & 7.97 & 8.24 & 8.33 & 7.83 \cite{Whited:1973eo} \\
  LiCl    &(RS)    & 6.54 & 8.66  & 9.42 & 9.57 & 9.62 & 9.4 \cite{Baldini:1970ba}  \\
  NaCl    &(RS)    & 5.18 & 7.26  & 8.55 & 8.73 & 8.84 & 8.6 \cite{Nakai:1969wu} \\
  LiF     &(RS)    & 9.21 & 12.28 & 15.48&15.83 &16.15 & 14.2  \cite{Piacentini:1976fw}  \\
  H$_2$O  &(XI)    & 5.57 & 8.05  &11.19 & 11.44& 11.71& 10.9 \cite{Kobayashi:1983wx}  \\
  Ar      &(cF)    & 8.78 & 11.20 & 14.40& 14.54& 14.67& 14.2 \cite{Schwentner:1975dp} \\
  Ne      &(cF)    &11.65 & 15.20 & 23.32& 22.99& 23.67& 21.7 \cite{Schwentner:1975dp}\\
%\cmidrule{3-8} 
\hline
  ME (eV)  && -2.7 & -0.3 & 0.0  & 0.3  &0.3   &   -- \\
  MAE (eV) && 2.67  &  1.08 & 0.5  & 0.4  &0.5   &   --   \\
  MRE (\%)  && -46.9  & 10.8 & -1.1 & 4.9  & 3.3  &   -- \\
  MARE (\%) &&  46.9  & 21.1& 9.6  & 7.4  & 7.8  &   --   \\
%\bottomrule
%\bottomrule
\hline
\hline
\end{tabular} }
\end{table*}
We now turn to the comparison of the Kohn-Sham gaps evaluated using the fixed dielectric-dependent hybrid functionals and the self-consistent dielectric-dependent functional (sc-hybrid) at the experimental geometries (Table~\ref{table:Eg-comp-tab}). We also included in Table~\ref{table:Eg-comp-tab} the results obtained with the GGA functional PBE and the fixed $\alpha=0.25$ hybrid functional PBE0. 
In most cases, we found a considerable improvement over GGA with hybrid functionals, with the best results obtained for the dielectric-dependent hybrid functionals. The largest relative errors were found for the insulators alumina (Al$_2$O$_3$) and hafnia (HfO$_2$). This discrepancy with experiments may be due, at least in part, to the poor crystallinity of the samples used experimentally. 
The presence of `band tail states' was investigated for hafnia, and a corrected photoemission gap, obtained by removing the band tails was reported (6.7 eV)\cite{Sayan:2004gx}, which is very similar to the one computed with the sc-hybrid functional (6.8 eV). The alumina experimental gap reported in Table~\ref{table:Eg-comp-tab} is an optical gap (excitonic contributions present). However, the exciton binding energy of alumina was estimated to be similar to that of excitons in MgO (0.06 eV),\cite{French:1990gz} and hence the optical and photoemission gaps are expected to differ at most by $\sim 0.1$ eV.

\begin{table}[] 
\caption{The Kohn-Sham (KS) energy gaps (eV) evaluated in the present work (see Table~\ref{table:Eg-comp-tab}) and quasiparticle gaps from Refs.~\onlinecite{Shishkin:2007ji,Shishkin:2007PRL} are compared with the measured electronic gaps for a subset of semiconductors and insulators. Both the sc$GW$ (RPA) and sc$GW$ (e-h) calculations used the HSE hybrid functional eigenvalues and orbitals as input, and the $G_{0}W_{0}$ (RPA) calculations used the PBE functional eigenvalues and orbitals as input. The sc-hybrid (NLF) heading refers to a hybrid calculation where the uncoupled-pertrubed Kohn-Sham equation was used, thus neglecting all local-field effects. The sc-hybrid (RPA $+ f_{xc-nl}$) heading refers to a sc-hybrid calculation where the polarizability includes all local-field effects. MRE and MARE stands for the mean relative and the mean absolute relative error, respectively. The experimental geometry was used for each solid, with the structure/polytype indicated in Table~\ref{table:Eg-comp-tab}.}
\footnotetext[2]{ denotes the experimental value used here is the average of two reported values 6.1 and 6.4 eV.}
\label{table:Eg-comp-Kresse}
\begin{adjustbox}{max width=\columnwidth,max height=\textheight,keepaspectratio}
\begin{tabular}{lcccccc}
%\toprule
%\toprule
\hline
\hline
               &\multicolumn{2}{c}{sc-hybrid } & $G_0W_0$ \cite{Shishkin:2007ji} &  sc$GW$\cite{Shishkin:2007PRL}& sc$GW$\cite{Shishkin:2007PRL}& Exp.  \\  
        & (NLF) & (RPA $+ f_{xc-nl}$) &  (RPA) & (RPA) & (e-h) & \\  
%\cmidrule{2-7} 
\hline
  Ge        &0.72 &0.71 &  --  & 0.95 & 0.81 & 0.74 \\
  Si        &1.00 &0.99 & 1.12 & 1.41 & 1.24 & 1.17 \\
  AlP       &2.42 &2.37 & 2.44 & 2.90 & 2.57 & 2.51 \\
  SiC       & 2.38&2.29 & 2.27 & 2.88 & 2.53 & 2.39 \\ 
  GaN       &3.47  &3.26 & 2.80 & 3.82 & 3.27 & 3.29 \\
  ZnO       &4.35 &3.78 & 2.12 & 3.8  & 3.2  & 3.44 \\
  ZnS       &3.96 &3.82 & 3.29 & 4.15 & 3.60 & 3.91 \\
  C         &5.55 &5.42 & 5.50 & 6.18 & 5.79 & 5.48 \\
  BN        &6.55 &6.33 & 6.10 & 7.14 & 6.59 & 6.25 \footnotemark[2] \\
  MgO       &8.93 &8.33 & 7.25 & 9.16 & 8.12 & 7.83 \\
  LiF       &-- &16.15 &13.27 &15.9  &14.5  &14.2  \\
  Ar        &14.88 &14.67 &13.28 &14.9  &13.9  &14.2  \\
  Ne        &24.05 &23.67 &19.59 &22.1  &21.4  &21.7  \\
%\cmidrule{2-7} 
\hline
  ME   (eV) &0.45 & 0.36 & -0.58&0.63&0.03& -- \\ 
  MAE  (eV) &0.49 & 0.46 & 0.58 &0.63&0.21& -- \\
  MRE  (\%) &4.0 &0.8  &-9.4 &13.9  &1.6   &  --   \\
  MARE (\%) &7.5 & 5.9  & 9.5 &13.9  &4.6   &  -- \\
%\bottomrule
%\bottomrule
\hline
\hline
\end{tabular}
\end{adjustbox}
\end{table}
Table ~\ref{table:Eg-comp-Kresse} compares the electronic gaps evaluated with the present sc-hybrid scheme and with the $GW$ approximation. The sc$GW$ ($G_0W_0$) calculations used the HSE (PBE) hybrid functional eigenvalues and wavefunctions as input. The error of the sc-hybrid functional in predicting band gaps is similar to the one introduced by the sc$GW$ method where e-h interactions are included in $W$.\\

We also computed the valence bandwidths for a subset of the solids listed in Table~\ref{table:epsilinf-comp-tab} and Table~\ref{table:Eg-comp-tab}; these are shown in Table~\ref{table:bandwidths-comp}. The results of the dielectric dependent hybrid functionals agree remarkably well with experiment, whereas the PBE and PBE0 functionals systematically underestimate and overestimate the bandwidths, respectively. There is an outlier, i.e. ZnO for which none of the computed valence bandwidths agree with experiment. 
\begin{table}[] 
\caption{The valence bandwidths (VBW, eV) are shown for a subset of the solids. For a description of the dielectric-dependent exact-exchange mixing scheme hybrid functional column headings see text and Table~\ref{table:Eg-comp-tab}.}
\footnotetext[3]{The value listed for SiC in the last column is the VBW obtained from $G_0W_0$ calculations using the plasmon-pole approximation and a model dielectric function (within IPA).}
\label{table:bandwidths-comp}
\begin{adjustbox}{max width=\columnwidth,max height=\textheight,keepaspectratio}
\begin{tabular}{lcccccccc}
%\toprule
%\toprule
\hline
\hline
               &  PBE & PBE0 & hybrid & hybrid &sc-hybrid & Exp. \\
&  $\alpha=0$  & $\alpha=0.25$  & $\alpha={1/\epsilon}^{\text{PBE}}_{{\infty} }$ & $\alpha={1/\epsilon}^{\text{PBE0}}_{{\infty} }$ & $\alpha=1/$sc-${\epsilon}_{\infty}$ &  \\  
%\cmidrule{2-7} 
\hline
  Si            &  11.9 & 13.4 & 12.4 & 12.5 & 12.4  & 12.5  \cite{Kevan:1992}\\
  C             &  13.4 & 23.6 & 23.0 & 23.0 & 23.0  & 23.0  \cite{Jimenez:1997wh}\\
  Ge            &  --   & 14.0 & --   & 13.0 & 12.9  & 12.9 \cite{Kevan:1992}\\
  SiC           &  15.4 & 17.0 & 16.3 & 16.4 & 16.4  & 16.9  \cite{Furthmuller:2002un}\footnotemark[3]\\
  LiF           &  3.1  &  3.3 & 3.3  & 3.4  & 3.4   & 3.5 \cite{Himpsel:1992}\\
  MgO           &  4.6  &  5.0 & 5.1  & 5.2  & 5.2   & 4.8 \cite{Tjeng:1990um}\\
  ZnO           &  6.1  &  7.0 & 6.7  & 7.0  & 7.2   & 9.0 \cite{Ozgur:2005it}\\ 
  TiO$_2$       &  5.7  &  6.4 & 6.1  & 6.2  & 6.1   & $\sim6.0$ \cite{Tezuka:1994cv}\\
%\bottomrule
%\bottomrule
\hline
\hline
\end{tabular}
\end{adjustbox}
\end{table}
Both hybrid density functionals, as well as $GW$, incorrectly describe the localized occupied d-band, with a tendency to underbind (see Table~\ref{table:dband-comp}).
Though sc$GW$ results are not shown in Table~\ref{table:dband-comp}, the authors of Ref.~\onlinecite{Shishkin:2007PRL} reported that the sc$GW$ band positions are underbound by a similar magnitude as the $G_0W_0$ results.
\begin{table*}[] 
\caption{The d band position relative to the valence band maximum (eV) is shown for a subset of solids with d electrons. The $G_0W_0$ and $GW_0$ results, as well as the experimental values are taken from Ref.~\onlinecite{Shishkin:2007ji}. For a description of the dielectric-dependent exact-exchange mixing scheme hybrid functionals column headings see Table~\ref{table:Eg-comp-tab}. }
\label{table:dband-comp}
\scalebox{1.0}{
\begin{tabular}{lcccccccc}
%\toprule
%\toprule
\hline
\hline
               &  PBE & PBE0 & hybrid & hybrid & sc-hybrid& $G_0W_0$ \cite{Shishkin:2007ji} &  $GW_0$\cite{Shishkin:2007ji}&  Exp.  \\  
&  $\alpha=0$  & $\alpha=0.25$  & $\alpha={1/\epsilon}^{\text{PBE}}_{{\infty} }$ & $\alpha={1/\epsilon}^{\text{PBE0}}_{{\infty} }$ & $\alpha=1/$sc-${\epsilon}_{\infty}$ & RPA & RPA &  \\  
%\cmidrule{2-9} 
\hline
  GaN             & -13.8 & -15.7& -16.0& -16.2& -16.6& -16.0&-16.9 & -17.0 \\
  ZnO             & -5.0  & -6.0 &  -5.9& -6.2 & -6.3 & -6.2 & -6.6 & -7.5,-8.81    \\
  ZnS             & -6.3  & -7.8 &  -7.5& -7.5 &  -7.5&  -7.0& -7.5 & -9.03  \\
%\bottomrule
%\bottomrule
\hline
\hline
\end{tabular} }
\end{table*}
%

%%%%%%%%%%%%%%%%%%%%%%%%%%%%%%%%%%%%%%
\subsection{{\bf Lattice constants }}
\begin{figure}[]
 \centering
  \scalebox{0.75}{ \includegraphics[width=0.65\textwidth,trim= 1mm 1mm 1mm 18mm,clip]{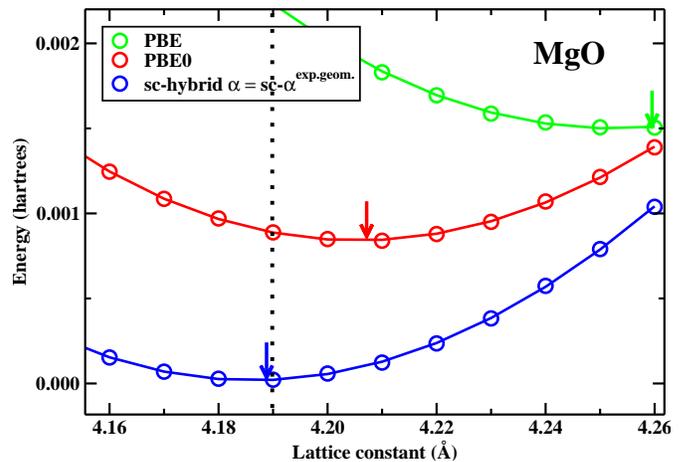}}
 \caption{(Color online) Total energy of MgO as a function of the lattice constant (\AA). The black dotted vertical line indicates the experimental value extrapolated to 0K where the zero-point anharmonic contribution has been removed. The colored arrows point to the minima of each surface. The PBE and PBE0 total energy curves were shifted by -0.027 and -0.006 hartrees, respectively, in order to fit on the same plot.}
\label{fig:opt-surfaces-MgO}
 \end{figure}

We further used the dielectric-dependent hybrid functionals to perform structural optimizations of a subset of solids. In most cases, including exact-exchange improves the agreement of the computed lattice constants with experiment for nonmetallic systems, as compared to the semilocal functional (PBE) results (see Table~\ref{table:opt-lat-constants}). 
\begin{table*}[] 
\caption{The equilibrium lattice constant (\AA) for a subset of solids compared with experiment. For a description of the dielectric-dependent exact exchange mixing scheme hybrid functionals column headings see Table~\ref{table:Eg-comp-tab}. The first column corresponds to the lattice constant evaluated with plane-wave (PW) basis and PAW pseudopotentials. All other results were obtained with Gaussian basis sets (GTO). The $\alpha$ fixed sc-hybrid column indicates the value of $\alpha$ is kept fixed throughout the lattice optimzation to the value of $\alpha$ deterimined self-consistently at the experimental geometry. The experimental lattice constants are from Ref.~\onlinecite{Schimka:2011cw}. The (0 K) column corresponds to the experimental measured lattice constant extrapolated to 0 Kelvin. The (ZPAE) column is obtained by removing from the experimental (0 K) column the zero-point anharmonic expansion effects, determined from first principles. Our calculated results should be compared with the experimental (ZPAE) column since in our calculations we do not include zero-point energy contributions.}
\label{table:opt-lat-constants}
\scalebox{1.0}{
\begin{tabular}{lcccccccccc}
%\toprule
%\toprule
\hline
\hline
               &  PBE PW & PBE GTO & PBE0 & hybrid & hybrid &sc-hybrid &\multicolumn{2}{c}{Exp.}  \\
 &  $\alpha=0$ &  $\alpha=0$ & $\alpha=0.25$  & $\alpha={1/\epsilon}^{\text{PBE}}_{{\infty} }$ & $\alpha={1/\epsilon}^{\text{PBE0}}_{{\infty} }$ & $\alpha$ fixed & (0 K)& (ZPAE)  \\  
%\cmidrule{2-9} 
\hline
  Si        &5.47& 5.47 & 5.44 & 5.46& 5.46&  5.46 &       5.43 & 5.42 \\ %done
  C         &3.57& 3.57 & 3.55 & 3.55& 3.55&  3.55 &       3.57 & 3.54 \\ %done
  SiC       &4.38& 4.38 & 4.35 & 4.37& 4.36&  4.36 &       4.36 & 4.34 \\ %done
  MgO       &4.26& 4.26 & 4.21 & 4.20& 4.19&  4.19 &       4.21 & 4.19 \\ %done
  LiCl      &5.15& 5.15 & 5.11 & 5.10& 5.10&  5.10 &       5.11 & 5.07 \\ %done
  NaCl      &5.69& 5.68 & 5.63 & 5.61& 5.61&  5.61 &       5.60 & 5.57 \\ %done
%\bottomrule
%\bottomrule
\hline
\hline
\end{tabular} }
\end{table*}
For the sc-hybrid functional, the total derivative of the energy $E\left(R,\alpha(R)\right)$ with respect to the lattice constant $R$, is expressed as: 
\begin{equation}
 \frac{dE}{dR} = {\left(\frac{\partial E}{\partial R}\right)} +  \left(\frac{\partial E}{\partial \alpha}\right)\frac{d\alpha}{dR}.
\label{equation:dEdR}
\end{equation}
%\begin{equation}
% \frac{dE}{dR}\bigg|_{R_0} = {\left(\frac{\partial E}{\partial R}\right)}_{\!\!\alpha_0} +  \left(\frac{\partial E}{\partial \alpha}\right)_{\!\!{R_0}}\frac{d\alpha}{dR}\bigg|_{R_0} = 0,
%\label{equation:dEdR}
%\end{equation}
%where $R_0$ is the equilibrium geometry and $\alpha_0$ is the self-consistent $\alpha$ evaluated at the equilibrium geometry $R_0$. 
When the second term on the right hand side of Eq.~\eqref{equation:dEdR} is much smaller than the first term, e.g. when $\alpha$ is almost constant as a function of $R$, close to the minimum, the total derivative of the energy can be approximated as: 
\begin{equation}
 \frac{dE}{dR} \cong {\left(\frac{\partial E}{\partial R}\right)}. 
\label{equation:dEdR-approx}
\end{equation}
The sc-hybrid lattice constants shown in Table~\ref{table:opt-lat-constants} and the sc-hybrid potential energy surface plotted for MgO in Fig.~\ref{fig:opt-surfaces-MgO} were evaluated using Eq.~\eqref{equation:dEdR-approx}.
We note that the derivative in Eq.~\eqref{equation:dEdR-approx} is to be evaluated at constant $\alpha$ and its root is nearly insensitive to which $\alpha$ is chosen, whether the one determined self-consistently at the experimental equilibrium positions or a parameter $\alpha$ computed for a lattice constant close to the experimental equilibrium. 
This can be seen for example by comparing the results obtained with PBE0, hybrid, and sc-hybrid functionals and shown in Table~\ref{table:opt-lat-constants}, which were obtained for different fixed values of $\alpha$, and yet yielded optimal lattice constants that differ by less than 0.02 \AA.

For most of the systems shown in Table~\ref{table:opt-lat-constants}, Eq.~\eqref{equation:dEdR-approx} is a good approximation to the total derivative. However in the case of NaCl and LiCl, the second term on the right hand side of Eq.~\eqref{equation:dEdR} is nonnegligible and the roots of Eq.~\eqref{equation:dEdR} and Eq.~\eqref{equation:dEdR-approx} are different. In this case the root of Eq.~\eqref{equation:dEdR} yields results in poor agreement with the experimental lattice constants (e.g. using Eq.~\eqref{equation:dEdR} we obtain 5.96 \AA $ $ for NaCl, and 5.35 \AA $ $ for LiCl).  

%Agreement with experiment is very good, of the same quality as with PBE0 (see for example the potential energy surfaces of PBE0 and sc-hybrid in Fig.~\ref{fig:opt-surfaces-MgO} for MgO). Allowing $\alpha$ to be relaxed during the geometry optimization, without considering its dependence on ionic positions, leads to similar lattice constants as with $\alpha$ fixed for the sp-semiconductors, however it leads to largely overestimated lattice constants for the ionic materials NaCl and LiCl. This overestimiation of lattice constants for ionic materials, when $\alpha$ is relaxed during the lattice optimization, neglecting its dependence on ionic positions, was also observed by the authors of Ref.~\onlinecite{Koller:2013ht}. The derivative of the total energy functional with respect to the ionic positions ($R_I$) contains the term $\frac{dE}{d\alpha}\frac{d\alpha}{dR_I}$. For the sp-semiconductors $\alpha$ does not vary substantially with respect to changes in ionic positions so that the contributions to the total energy derivative arising from the derivative of $\alpha$ can be neglected, but this is not the case for the ionic materials where $\alpha$ has a non-negligible change with respect to ionic positions.  

%
%
%%%%%%%%%%%%%%%%%%%%%%% EOF %%%%%%%%%%%%%%%%%%%%%%%%%%%%%%%%
%
%%%%%%%%%%%%%%%%%%%%%%%%%%%%%%%%%%%%%%%%%%%%%%%%%%%%%%%%%%%%%%%%%%%%%%%%%%%%%%%%%%%%%%%%%%%%%%%%%%%%%%%%%%%%%%%%%%%%%%%%%%%%%%%%%%%%%
%%%%%%%%%%%%%%%%%%%%%%%%%%%%%%%%%%%%%%%%%%%%%%%%% SUMMARY AND CONCLUSIONS  %%%%%%%%%%%%%%%%%%%%%%%%%%%%%%%%%%%%%%%%%%%%%%%%%%%%%%%%%%
%%%%%%%%%%%%%%%%%%%%%%%%%%%%%%%%%%%%%%%%%%%%%%%%%%%%%%%%%%%%%%%%%%%%%%%%%%%%%%%%%%%%%%%%%%%%%%%%%%%%%%%%%%%%%%%%%%%%%%%%%%%%%%%%%%%%%
\section{ \textbf {Summary and Conclusions}}
We presented a full-range hybrid functional for the calculation of the electronic properties of nonmetallic condensed systems, which yielded results in excellent agreement with experiments for the band gaps and dielectric constants of a wide range of semiconductors and insulators. The exchange-correlation functional is defined in a way similar to the PBE0 functional, but the mixing parameter is set equal to the inverse macroscopic dielectric constant and it is determined self-consistently, by computing the optimal dielectric screening. 
We showed that convergence is usually achieved in 3 to 4 iterations, regardless of whether the initial value of  the dielectric constant is computed  at the PBE or PBE0 level of theory. In many cases the results for $\alpha = 1/\epsilon^{\textrm{PBE0}}_{\infty}$ are of similar accuracy to the sc-hybrid results,\footnote{Notable exceptions are CoO and NiO for which the sc-hybrid and the $\alpha = 1/\epsilon^{\textrm{PBE0}}_{\infty}$ hybrid were not as similar because the convergence in the sc-hybrid is not achieved till 5 and 9 iterations, respectively, for CoO and NiO.} which suggests that for certain systems self-consistency may be avoided; further reducing computational cost.
The presence of $f_{xc}$ in the local fields was investigated in detail, with particular emphasis on the nonlocal exchange contribution $f_{x-nl}$, which yields an accurate description of the static dielectric constant, when included. Our results suggest that including the nonlocal contributions in $f_{xc}$ is an effective way of including long-range interaction effects in condensed phase systems, without resorting to expensive vertex corrections.   
The computed band gaps and dielectric constants are in general much improved with respect to those obtained with the PBE and PBE0 functionals, with errors with respect to experiments similar in magnitude to those of fully self-consistent $GW$ (e-h) calculations. 

All results presented here were obtained within an all electron scheme (except for W and Hf for which we used effective-core pseudopotentials) and using first order perturbation theory within the CPKS scheme to compute the dielectric constant. Work is in progress to implement finite field methods for the dielectric constant in plane-wave pseudopotential codes, which will allow for the use of the sc-hybrid scheme for liquids, and in general disordered systems and  in ab-initio molecular dynamics calculations.
Though here we chose a full-range hybrid functional, our approach may be easily extended to range-separated hybrid functionals, where the static dielectric constant is used to define the mixing parameter of the long-range component. The computational cost of the sc-hybrid scheme is similar to that of hybrid calculations, making it a computationally cheaper alternative to $GW$ calculations. We note that a self-consistent dielectric screened hybrid functional provides a means to compute an effective statically screened Coulomb interaction $W$, and thus it offers a suitable starting point for $G_0W_0$ calculations.

\section*{ \textbf {Acknowledgments}}
We thank Michel R\'erart and Roberto Orlando for helpful discussions pertaining to the CPKS implementation in CRYSTAL. We also wish to thank Ding Pan for helpful discussions and for providing the ice XI structure. This work was supported by the NSF Center for Chemical Innovation (Powering the Planet, grant number NSF-CHE-0802907) and by the Army Research Laboratory Collaborative Research Alliance in Multiscale Multidisciplinary Modeling of Electronic Materials (CRL-MSME, grant number W911NF-12-2-0023). All calculations were performed at the NAVY DoD Supercomputing Resource Center of the Department of Defense High Performance Computing Modernization Program.
%%%%%%%%%%%%%%%%%%%%%%%%%%%%%%%%%%%%%%%%%%%%%%%%%%%%%%%%%%%%%%%%%%%%%%%%%%%%%%%%%%%%%%%%%%%%%%%%%%%%%%%%%%%%%%%%%%%%%%%%%%%%%%%%%%%%%
%%%%%%%%%%%%%%%%%%%%%%%%%%%%%%%%%%%%%%%%%%%%%%%%% SUPPLEMENTAL INFORMATION %%%%%%%%%%%%%%%%%%%%%%%%%%%%%%%%%%%%%%%%%%%%%%%%%%%%%%%%%%
%%%%%%%%%%%%%%%%%%%%%%%%%%%%%%%%%%%%%%%%%%%%%%%%%%%%%%%%%%%%%%%%%%%%%%%%%%%%%%%%%%%%%%%%%%%%%%%%%%%%%%%%%%%%%%%%%%%%%%%%%%%%%%%%%%%%%

%\section*{ \bf {Supplemental Information}}

%%%%%%%%%%%%%%%%%%%%%%%%%%%%%%%%%%%%%%%%%%%%%%%%%%%%%%%%%%%%%%%%%%%%%%%%%%%%%%%%%%%%%%%%%%%%%%%%%%%%%%%%%%%%%%%%%%%%%%%%%%%%%%%%%%%%%
%%%%%%%%%%%%%%%%%%%%%%%%%%%%%%%%%%%%%%%%%%%%%%%%%%%%%%% REFERENCES  %%%%%%%%%%%%%%%%%%%%%%%%%%%%%%%%%%%%%%%%%%%%%%%%%%%%%%%%%%%%%%%%%
%%%%%%%%%%%%%%%%%%%%%%%%%%%%%%%%%%%%%%%%%%%%%%%%%%%%%%%%%%%%%%%%%%%%%%%%%%%%%%%%%%%%%%%%%%%%%%%%%%%%%%%%%%%%%%%%%%%%%%%%%%%%%%%%%%%%%
\bibliography{references-scHybrid}
\label{LastPage}
%\bibliographystyle{FG-bibstyle}

%%%%%%%%%%%%%%%%%%%%%%%%%%%%%%%%%%%%%%%%%%%%%%%%%%%%%%%%%%%%%%%%%%%%%%%%%%%%%%%%%%%%%%%%%%%%%%%%%%%%%%%%%%%%%%%%%%%%%%%%%%%%%%%%%%%%%
%%%%%%%%%%%%%%%%%%%%%%%%%%%%%%%%%%%%%%%%%%%%%%%%%%%%%%%%% END OF DOC %%%%%%%%%%%%%%%%%%%%%%%%%%%%%%%%%%%%%%%%%%%%%%%%%%%%%%%%%%%%%%%%
%%%%%%%%%%%%%%%%%%%%%%%%%%%%%%%%%%%%%%%%%%%%%%%%%%%%%%%%%%%%%%%%%%%%%%%%%%%%%%%%%%%%%%%%%%%%%%%%%%%%%%%%%%%%%%%%%%%%%%%%%%%%%%%%%%%%%
\end{document}